\documentclass[pra,amsmath,amsfonts,11pt,letterpaper,nofootinbib,showpacs,showkeys]{revtex4}
\usepackage{epsfig,amsmath,amssymb,bm,epsf,graphics,latexsym,amsfonts}

\def\ket#1{ | #1 \rangle}
\def\bra#1{{\langle #1 | }}
\def\tr{ {\rm{Tr }}}
\newcommand{\avg}[1]{\langle #1 \rangle}

\begin{document}

\title{ On Decoherence in Quantum Clock Synchronization}

\author{Sergio Boixo}
\affiliation{Department of Physics and Astronomy, University of New Mexico, Albuquerque, New Mexico 87131-1156, USA.}
\author{Carlton M. Caves}
\affiliation{Department of Physics and Astronomy, University of New Mexico, Albuquerque, New Mexico 87131-1156, USA.}
\author{Animesh Datta}
\email{animesh@unm.edu}
 \affiliation{Department of Physics and Astronomy, University of New Mexico, Albuquerque, New Mexico 87131-1156, USA.}
\author{Anil Shaji}
\affiliation{Department of Physics and Astronomy, University of New Mexico, Albuquerque, New Mexico 87131-1156, USA.}

\date{\today}

\begin{abstract}

We study two quantum versions of the Eddington clock-synchronization
protocol in the presence of decoherence.  The first protocol uses
maximally entangled states to achieve the Heisenberg limit for clock
synchronization.  The second protocol achieves the limit without using
entanglement.  We show the equivalence of the two protocols under any
single-qubit decoherence model that does not itself provide
synchronization information.

\end{abstract}

\maketitle
\section{Introduction}
\label{intro}

The problem of synchronizing distant clocks is of fundamental interest in
physics, as well as having important applications in metrology and
engineering.  Suppose Alice and Bob are two space-like-separated observers
who wish to synchronize their clocks.  We consider the situation in which
the two observers share an inertial reference frame and have two classical
clocks ticking at the same rate.  Classically, there are two canonical
protocols for synchronizing clocks, one due to Einstein, which is based on
sending light signals, and one due to Eddington, which involves exchanging
clocks.  The accuracy with which Alice and Bob can synchronize their
clocks classically, using either procedure, scales as $1/\sqrt n$, where
$n$ is the number of times the protocol is executed. This scaling is generally 
known as the Standard Quantum Limit~(SQL) \cite{braginsky75a,caves80a,giovannetti04a}.

Over the last decade, there has been considerable interest in studying
quantum versions of these protocols
\cite{bollinger96a,huelga97a,jozsa00a,chuang00a, preskill00a,
yurtsever02a,revzen03a,giovannetti04a,burgh05a}. Some of this work was
motivated solely by the quest for better frequency standards, whereas
others aimed at beating the SQL.  It has now been shown that quantum
clock-synchronization protocols can perform better than classical ones and
that the scaling can be improved in the quantum case to $1/n$, the
so-called Heisenberg Limit.

There are two interesting quantum versions of the Eddington protocol for
clock synchronization, which classically involves Alice (adiabatically)
sending to Bob a ``watch'' synchronized with her clock.  In quantum
versions of this protocol, the watch is a time evolving observable of one
or more qubits.  In one version, these ``ticking qubits'' are prepared in
a maximally entangled ``cat'' state~\cite{bollinger96a,jozsa00a};
entanglement is the resource that allows this protocol to achieve the
Heisenberg Limit. An alternative quantum version of the Eddington protocol
achieves the Heisenberg Limit\footnote{The authors of Ref.~\cite{burgh05a}
argue that the fundamental limit for clock synchronization scales as $\ln
n/n$.} by using multiple, coherent exchanges of a single ticking qubit
between Alice and Bob~\cite{burgh05a}; in this protocol quantum coherence
is identified as the resource that provides an advantage over the
classical protocol.

Our aim in this paper is to compare and contrast the performance of the
two protocols in the presence of decoherence in the quantum channel
between Alice and Bob.  It is known that both quantum versions of the
Eddington protocol use the same amount of total qubit communication
\cite{burgh05a}.  Our study investigates the status of this similarity in
the case of a non ideal quantum channel between Alice and Bob, which is
expected to degrade the performance of the protocols.  We find that the
two protocols are affected in the same way by any single-qubit decoherence
process that does not itself provide useful synchronization information.

The paper is structured as follows. In Sec.~\ref{sec2} we review the two
quantum clock-synchronization protocols and touch on the question of
whether anything more than a shared inertial reference frame is required
for Alice and Bob to synchronize their clocks.  Section~\ref{decoherence}
describes our decoherence model.  Our results are presented in
Sec.~\ref{sec4}, and we conclude in Sec.~\ref{sec5}.

\section{Two Quantum Clock-Synchronization Protocols}
\label{sec2}

\subsection{Assumptions and conventions}
\label{sec2A}

Prior to describing clock-synchronization protocols, we need to describe
our assumptions and fix some conventions.  The problem is to synchronize
two classical clocks, one maintained by Alice, which reads time $t_A$, and
one by Bob, which reads time $t_B$.  We assume that Alice's and Bob's
clocks tick at the same rate, so the problem is wholly that of determining
the constant offset $t_{BA}=t_B-t_A$.

Since we are considering quantum versions of the Eddington protocol, Alice
and Bob synchronize their clocks by exchanging ticking qubits.  We think
of a ticking qubit as being two atomic levels, whose free Hamiltonian is
\begin{equation}
H_0={\frac{\hbar\omega}{2}}Z=
{\frac{\hbar\omega}{2}}\bigl(|0\rangle\langle0|-|1\rangle\langle1|\bigr)\;,
\end{equation}
where $Z$ is the Pauli $\sigma_z$ operator and $|0\rangle$ and $|1\rangle$
are the upper and lower energy eigenstates.  Agreeing on the atomic free
Hamiltonian means, first, that Alice and Bob agree on the $Z$ axis of the
Bloch sphere that describes the two-dimensional atomic Hilbert space and,
second, that they share the transition frequency $\omega$.  This frequency
is an expression of the fixed frequency unit that the parties share as a
consequence of having clocks that tick at the same rate.

To exchange information through the ticking qubits, Alice and Bob must
perform operations on the qubits, such as setting them ticking, stopping
their ticking, and changing the phase of their ticking.  All these
operations can be performed by illuminating a qubit with a laser tuned to
the transition frequency.  The Hamiltonian for this interaction, in the
interaction picture, is the Rabi Hamiltonian
 \begin{equation}
 \label{eq:ham}
H_{\rm Rabi}=
\frac{\hbar\Omega}{2}\bigl(e^{-i\varphi_P}\ket{0}\bra{1}
+e^{i\varphi_P}\ket{1}\bra{0}\bigr)
=\frac{\hbar\Omega}{2}\bigl(X\cos\varphi_P+Y\sin\varphi_P)\;.
 \end{equation}
Here $X$ and $Y$ denote the Pauli $\sigma_x$ and $\sigma_y$
operators, $\Omega$ is the Rabi frequency, and $\varphi_P$ is the phase of
the driving laser relative to the zero of clock~$P$.  In Bloch-sphere
language, the Rabi Hamiltonian generates a rotation at frequency $\Omega$
about an axis in the equatorial plane that makes an angle $\varphi_P$ with
the $X$ axis.

Since Alice and Bob do not share clocks with a common zero, they have
different phase references $\varphi_A$ and $\varphi_B$ and, hence,
different Bloch-sphere $X$ and $Y$ axes.  The problem of clock
synchronization reduces to estimating the phase offset
$\varphi_{BA}=\varphi_B-\varphi_A=\omega t_{BA}\equiv\varphi$.  When
$\varphi\neq0$, Alice and Bob describe states and operations
differently.  Their separate descriptions are related by a rotation
through angle $\varphi$ about the common $Z$ axis~\cite{burgh05a}. An
operator $O_A$ in Alice's description is considered by Bob to be the
operator
 \begin{equation}
 \label{eq:intro2}
 O_B= e^{-iZ\varphi/2}O_A e^{iZ\varphi/2}\;.
 \end{equation}

It might seem that the sharing of a common Bloch-sphere $Z$ axis requires
a preferred spatial direction shared by the two parties.  It is, however,
possible to imagine a situation in which both $\ket{0}$ and $\ket{1}$ are
zero-angular-momentum levels of an atom, in which case the
Hamiltonian~(\ref{eq:ham}) has no preferred spatial direction. Unfortunately,
electric-dipole ($E1$) selection rules forbid any $J'=0\rightarrow J=0$
transition.  To circumvent this, one could use a coherent two-photon Raman
process via an intermediate state to drive the forbidden transition.
Although this process does involve fixed orientations in space, since the
Raman transitions are carried out by Alice and Bob independently and
locally in their respective laboratories, they need not share a
\emph{common\/} spatial axis.  We thus do \emph{not\/} need the alignment
of any spatial axes for time synchronization.  This is in harmony with the
process of aligning spatial reference frames, which does not require any
time synchronization~\cite{rudolph03a}.

\subsection{Cat-state entangled protocol}
\label{sec2B}

Of the schemes designed to beat the SQL, one of the earliest uses the
entangled ``cat'' state of $n$ qubits \cite{bollinger96a}.  Starting with
the state $\ket{000\ldots 0}$, Alice prepares the cat
state~\cite{huelga97a},
 \begin{equation}
\ket{\psi_A} = \frac{1}{\sqrt{2}}\bigl(\ket{000\ldots0} +
\ket{111\ldots1}\bigr)_A\;,
 \end{equation}
by performing a $90^\circ$ rotation about the $Y_A$ axis (or a Hadamard
gate $H_A$) on the first qubit, followed by controlled spin flips ($X_A$),
controlled by the first qubit and targeted on each of the other qubits.
Alice sends the $n$ qubits to Bob, who measures the observable
$O_B=X_B^{\otimes n}$ (alternatively, Bob can reverse the steps that
prepared the cat state, but using his operations, of course, and then
measure $Z$ on the first qubit).  Since $O_B$ is a binary observable, its
distribution is determined by its expectation value,
\begin{equation}
\bra{\psi_A}O_B\ket{\psi_A}=\cos(n\phi_{BA})\;,
\end{equation}
which corresponds to Ramsey-fringe probabilities
 \begin{equation}
 \label{eq:fringes}
p_{\pm}=[1\pm\cos(n\phi_{BA})]/2
 \end{equation}
for results $\pm1$.  This leads to a nominal uncertainty in determining
$\varphi$ given by
 \begin{equation}
 \label{eq:resolution}
 \Delta\varphi=
 \frac{\Delta O_B}{|d\langle O_B\rangle/d\varphi|}=\frac{1}{n}\;,
\end{equation}
where $\Delta O_B=\sin(n\varphi)$ is the uncertainty in $O_B$.

Since the probabilities~(\ref{eq:fringes}) are periodic in $\varphi_{BA}$
with fringe period $2\pi/n$, determining $\varphi$ within the
uncertainty~(\ref{eq:resolution}) requires one already to know $\varphi$
to an accuracy $2\pi/n$.  One gets around this problem by using an
extended protocol, which ultimately leads to a Heisenberg-limited
sensitivity.  Defining $\omega t_{BA}=\varphi=\pi T$ and writing the
dimensionless time offset $T$ in binary form as $T=0.t_1t_2\ldots\,$, the
problem of synchronizing clocks becomes that of determining the sequence
of bits in the binary decomposition.  (We are assuming that Alice and Bob
already know $\varphi$ to within $\pi$, but notice that Bob could
determine the bit $t_0$ in $T=t_0.t_1t_2\ldots$ by running the bare
protocol several times with a single unentangled qubit).  If the sequence
is known through the $(j-1)$th bit, ascertaining the $j$th bit can be
accomplished by using the cat state with $2^j$ qubits.  Of course, Alice
and Bob must repeat the bare protocol several times to build up the
statistics to determine the $j$th bit.  The statistical uncertainty in
$\varphi$, given by $1/(2^j\sqrt\nu)$, where $\nu$ is the number of
repetitions, should be small compared to $\pi/2^j$; i.e., $\pi\sqrt\nu$
should be somewhat larger than 1 in order to determine the $j$th bit
reliably.

Our conclusion is that to determine the first $k\gg1$ bits of $T$ requires
running the bare protocol $\nu$ times for each bit, using $2^j$ entangled
qubits to determine the $j$th bit, for a total of $N=2\nu(2^k-1)$ qubits.
The resulting accuracy in determining $t_{BA}$,
 \begin{equation}
 \label{eq:resolution2}
\delta t_{BA}=\frac{1}{\omega}\frac{\pi}{2^k}\simeq\frac{2 \pi \nu}{\omega N}\;,
 \end{equation}
has the scaling of the Heisenberg limit.

\subsection{Coherent-transport protocol}
\label{sec2C}

\begin{figure}[!b]
\resizebox{6 in}{1.5 in}{\includegraphics{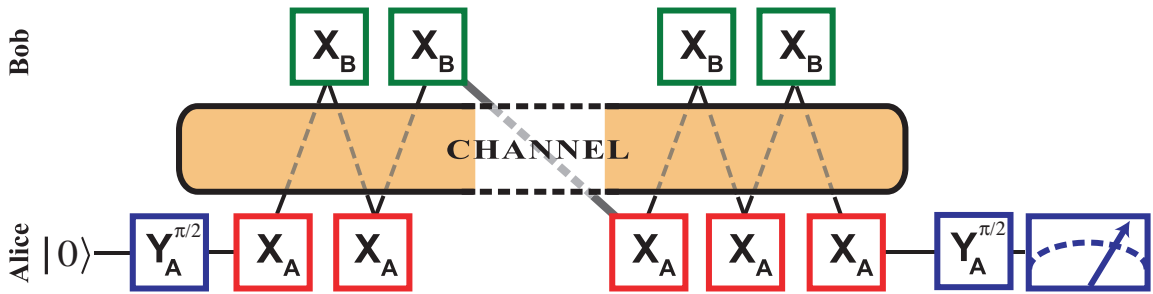}}
\caption{(Color online) The coherent-transport protocol in pictures. The
$90^\circ$ rotation about $Y_A$, denoted here by $Y_A^{\pi/2}$, takes
state $\ket{0}$ to $(\ket{0}+\ket{1})_A/\sqrt2$; $X_A$ and $X_B$ are
$180^\circ$ rotations by Alice \and Bob about their respective $X$ axes. A
final $90^\circ$ rotation by Alice is followed by a measurement of $Z$. An
alternative end point for the protocol is for Alice to return the qubit one
last time to Bob, who does a $90^\circ$ rotation about his axis $Y_B$,
followed by a measurement of $Z$.}
\label{fig1}
\end{figure}

The procedure outlined in the preceding subsection uses entanglement of a
larger and larger number of qubits to read out successive digits of $T$,
so we call it the \emph{entanglement protocol}.  The use of entanglement
is in line with the notion that entanglement is necessary to beat the SQL.
There exist protocols, however, that do not require entanglement to beat
the SQL, relying instead on coherent exchanges of a single qubit. The first such 
protocol was presented by Rudolph and Grover~\cite{rudolph03a} for the task of 
aligning spatial reference axes. The underlying idea of multiple exchanges was 
considered much earlier, in a wider context, by Salecker and Wigner~\cite{salecker58a}.  
Rudolph and Grover's protocol was adapted to the problem of synchronizing clocks 
by de Burgh and Bartlett~\cite{burgh05a}.

In this protocol Alice prepares a qubit in the state $\ket{0}$ and applies
her $90^\circ$ rotation about $Y_A$ (or her Hadamard gate $H_A$) to put
the qubit in the state
 \begin{equation}
 \label{eq:phiA}
 |\phi_A\rangle=(|0\rangle+|1\rangle)_A/\sqrt2\;.
 \end{equation}
Alice then sends the qubit to Bob, who performs his operation $X_B$.  Bob
sends the qubit back to Alice, and she performs her operation $X_A$.  The
result of this exchange is that they jointly execute the operation
 \begin{equation}
\label{eq:intro6}
 X_A X_B = X_A (e^{-iZ\varphi/2}X_A e^{iZ\varphi/2})=
 e^{iZ\varphi}\;.
 \end{equation}
Alice and Bob continue ping-ponging the qubit in this way.  If, after $r$
such exchanges, Alice measures the observable $O_A=X_A$ (alternatively,
she could undo the initial rotation about $Y_A$ and then measure $Z$), the
expectation value of $O_A$ is $\langle\phi_A|e^{-iZr\varphi}X_A
e^{iZr\varphi}|\phi_A\rangle=\cos(2\,r\varphi)$.  If, instead,
Alice returns the qubit to Bob, who measures $O_B=X_B$ (alternatively, Bob
could undo the initial $Y$ rotation, using his axis $Y_B$, of course, and
then measure $Z$), the expectation value of the measured observable is
$\langle\phi_A|e^{-iZr\varphi}X_B
e^{iZr\varphi}|\phi_A\rangle=\cos[(2\,r+1)\varphi]$.

We call this the \emph{coherent-transport protocol\/} because the qubit is
shuttled coherently back and forth between Alice and Bob.  The number of
uses of the qubit channel, $n=2r$ or $n=2r+1$, plays exactly the same role
in this protocol as does the number of qubits in the entangled protocol.
Indeed, since each qubit in the entangled protocol traverses the qubit
channel once, $n$ denotes the number of uses of the qubit channel for both
protocols.

The coherent-transport protocol can be generalized to an extended protocol
which reads out successive bits of the dimensionless time offset $T$, in
precise analogy to the extended entangled protocol.  In the extended
protocol, to determine the $j$th bit of $T$, Alice and Bob exchange the
qubit $r=n/2=2^{j-1}$ times, running the bare protocol several times to
build up sufficient statistics.  The coherent-transport protocol achieves
exactly the same sensitivity~(\ref{eq:resolution2}) as the entangled
protocol, with $n$ being the total number of uses of the quantum channel.
Notice that in the extended protocol, Alice makes all the measurements
and ends up with the measured value of $T$.

\begin{figure}[!ht]
\resizebox{6 in}{1.5 in}{\includegraphics{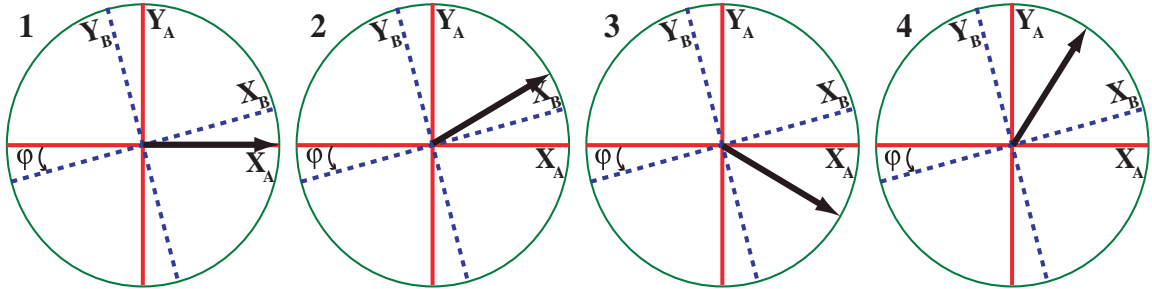}}
\caption{(Color online) How the coherent-transport protocol works.  In 1,
the Bloch vector (black arrow) of the qubit is shown after Alice has done
the initial rotation $Y^{\pi/2}_A$, which leaves the qubit polarized along
Alice's axis $X_A$.  Bob's axes $X_B$ and $Y_B$ (dashed blue lines) are
oriented at an angle $\varphi$ relative to Alice's axes (solid red
lines).  The qubit is sent to Bob who does a $180^\circ$ rotation about
$X_B$.  After this rotation, the Bloch vector, depicted in 2, has rotated
by an angle $2\varphi$ relative to $X_A$.  The Bloch vector after
Alice has done her $X_A$ operation is shown in 3, while in 4, the Bloch
vector after a further $X_B$ operation is shown. We see that each time
Alice and Bob send the qubit back and forth once, the Bloch vector is
rotated by angle $-2\varphi$. The original angle between the two sets
of axes is amplified and recorded in the state of the qubit when it is
exchanged several times between Alice and Bob.}
\label{fig2}
\end{figure}

The entangled protocol relies on entanglement to beat the SQL, whereas the
coherent-transport protocol relies on maintaining the coherence as it is
shuttled back and forth between Alice and Bob.  The former requires
maintaining spatial coherence among many qubits, whereas the latter
requires maintaining the temporal coherence of a single qubit.  Both
schemes are vulnerable to decoherence in the quantum channel between Alice
and Bob.  For the entangled protocol, calculations done using a specific
decoherence model revealed the deleterious effects of
decoherence and showed that other initial entangled states perform better 
than the cat state~\cite{huelga97a}.  The effect of decoherence on estimating the
time difference is, not surprisingly, related to the question of the
statistical distinguishability of neighboring states \cite{braunstein94a}.
The performance of the coherence-transport protocol also deteriorates in
the presence of decoherence in the channel.  Both protocols are able to
beat the SQL in the presence of a decoherence in the channel, albeit only
up to a limited precision governed by the level of noise in the channel.

Up till now, there has been no systematic study of the two quantum
clock-synchronization protocols in the presence of a general decoherence
model.  We provide such an analysis in the next section, where we consider
the most general single-qubit decoherence process possible for the
scenario at hand and study its effect on the two protocols.  The study
makes evident that the two protocols are essentially equivalent in their
sensitivity to decoherence.  The one difference that emerges prompts us to
propose a variation of the entangled protocol, which makes it precisely
equivalent to the coherent-transport protocol in the presence of
decoherence.

\section{Decoherence Model}\label{decoherence}

We model decoherence in the quantum channel as a completely positive,
trace-preserving (CPTP) linear map or superoperator~\cite[Chapter 8]{nielsen00a}, 
which acts on a qubit each time it traverses the quantum
channel.  This means that we ignore possible spatially or temporally
correlated decoherence that may occur in the two clock-synchronization
protocols.  A CPTP map $\mathcal{E}$ acting on a single qubit is defined
in terms of its action on the operator basis set
$\{\sigma_\xi\}=\{I,Z,X,Y\}$, i.e.,
 \begin{equation}
 \label{eq:matrix}
 \mathcal{E}(\sigma_\xi)=\sum_\eta\sigma_\eta\mathcal{E}_{\eta\xi}\;.
 \end{equation}
In writing this representation, we assume that $X$ and $Y$ are as defined
by Alice.  The $4 \times 4$ matrix that represents $\mathcal{E}$ has the
general form~\cite[Chapter 8]{nielsen00a}
 \begin{equation}
 \label{eq:E}
||\mathcal{E}_{\eta\xi}||=
\left(
\begin{array}{cccc}
1   & 0                 & 0                 & 0                 \\
t_Z &\phantom{R_2SR_1}  &\phantom{R_2SR_1}  &\phantom{R_2SR_1}  \\
t_X &                   & R_2SR_1           &                   \\
t_Y &                   &                   &
\end{array}
\right)\;,
 \end{equation}
where $R_1$ and $R_2$ are three-dimensional Bloch rotation matrices and
$S$ is a three-dimensional diagonal matrix,
 \begin{equation}
 S=
\left(
\begin{array}{ccc}
s_Z    &0      &0      \\
0      &s_X    &0      \\
0      &0      &s_Y
\end{array}
\right)\;,
 \end{equation}
whose diagonal elements satisfy $|s_j|\le1$.

We can define a related operation $\mathcal{F}$ whose matrix representation
is
 \begin{equation}
 \label{eq:F}
||\mathcal{F}_{\eta\xi}||=
\left(
\begin{array}{cccc}
1   & 0     & 0     & 0     \\
t_Z & s_Z   & 0     & 0     \\
t_X & 0     & s_X   & 0     \\
t_Y & 0     & 0     & s_Y
\end{array}
\right)\;.
 \end{equation}
The action of $\mathcal{E}$ is that of $\mathcal{F}$ preceded by rotation
$R_1$ and succeeded by rotation $R_2$, i.e.,
 \begin{equation}
\mathcal{E}(\rho)=U_2\mathcal{F}\bigl(U_1\rho U_1^\dagger\bigr)U_2^\dagger\;,
 \end{equation}
where $U_1$ and $U_2$ are the unitary operators corresponding to the
Bloch rotations $R_1$ and $R_2$.

In the interaction picture, clock synchronization reduces to finding the
angle $\varphi$ between Alice's axis $X_A$ and Bob's axis $X_B$.  We
assume that the channel itself, through the decoherence it produces,
should not provide any information about $\varphi$.  Formally, this
means that the map $\mathcal{E}$ should commute with rotations about the
common $Z$ axis.  This implies, first, that $t_X=t_Y=0$ (we let $t_Z=t$)
and, second, that the pre- and post-rotations $R_1$ and $R_2$ must be
rotations about the $Z$ axis and $s_X=s_Y=\lambda$ (we let $s_Z=s$).  With
these restrictions, $R_1$ and $R_2$ commute with $S$, so we can combine
them into a single (pre- or post-) rotation $R=R_2R_1$ about the $Z$ axis,
which we take to be a rotation by angle $\alpha$.  The
matrix~(\ref{eq:matrix}) of our CPTP map now takes the form
 \begin{equation}
 \label{eq:deco4}
 ||\mathcal{E}_{\eta\xi}||
 =
 \left(
\begin{array}{cccc}
1 & 0 & 0 & 0 \\ t & s & 0 & 0 \\
0 & 0 & \lambda\cos\alpha & -\lambda\sin\alpha \\
0 & 0 & \lambda\sin\alpha & \lambda\cos\alpha
\end{array}
\right)\;.
 \end{equation}

The matrix of the related operation $\mathcal{F}$ is even simpler,
 \begin{equation}
 \label{eq:F2}
 ||\mathcal{F}_{\eta\xi}||
 =
 \left(
\begin{array}{cccc}
1 & 0 & 0 & 0 \\ t & s & 0 & 0 \\
0 & 0 & \lambda & 0 \\
0 & 0 & 0 & \lambda
\end{array}
\right)\;,
 \end{equation}
and corresponds to a displacement of the Bloch sphere by a distance $t$ in
the $Z$ direction, compression of the Bloch sphere by a factor $s$ along
the $Z$ axis, and compression by a factor $\lambda$ in the equatorial
plane.  The operation of $\mathcal{E}$ is that of $\mathcal{F}$ with a
preceding or succeeding rotation by $\alpha$ about the $Z$ axis, i.e.,
 \begin{equation}
\mathcal{E}(\rho)=e^{-iZ\alpha/2}\mathcal{F}(\rho)e^{iZ\alpha/2}=
\mathcal{F}\bigl(e^{-iZ\alpha/2}\rho e^{iZ\alpha/2}\bigr)
\;.
 \end{equation}

The channel decoherence acts separately in the operator subspace spanned
by $I$ and $Z$ and the subspace spanned by $X$ and $Y$.  In the two
clock-synchronization protocols described in Sec.~\ref{sec2}, the last
step is a measurement by Alice or Bob of an operator in the equatorial
plane of the Bloch sphere.  As a result, we are only interested in the
part of the output density operator that lies in the $X$-$Y$ subspace.
Because $\mathcal{E}$ acts separately in the $I$-$Z$ and $X$-$Y$
subspaces, this means we only need to consider the part of $\mathcal{E}$
that acts in the $X$-$Y$ subspace.

Formally, we deal with this by introducing a superoperator projector $\Pi$
that projects any operator into the operator subspace spanned by $X$ and
$Y$:
 \begin{equation}
 \Pi(O) = \Pi(a_0 I + a_Z Z + a_X X + a_Y Y) = a_X X + a_Y Y\;.
 \end{equation}
Notice that it does not matter whether we use Alice's or Bob's $X$ and $Y$
operators to define $\Pi$.  The action of $\Pi$ can be written in two
other useful forms:
 \begin{equation}
\Pi(O)=\frac{1}{2}(O-ZOZ)=
|0\rangle\langle1|\langle0|O|1\rangle+|1\rangle\langle0|\langle1|O|0\rangle
\;.
 \end{equation}
The effect of $\Pi$ is to remove the diagonal matrix elements of $O$ in
the $Z$ basis, leaving the off-diagonal matrix elements.  The map is
neither trace preserving nor completely positive.

For our purposes, the crucial property of $\Pi$ is that it is Hermitian
relative to the operator inner product, i.e.,
 \begin{equation}
 \label{eq:Pihermiticity}
\tr\bigl(N^\dagger\Pi(O)\bigr)=\tr\Bigl(\bigl(\Pi(N)\bigr)^\dagger O\Bigr)\;.
 \end{equation}
It is also easy to see that $\Pi\circ\mathcal{F}(O)=\lambda\,\Pi(O)$,
from which it follows that
 \begin{equation}
 \label{eq:proj}
\Pi\circ\mathcal{E}(O)=
\lambda\,e^{-iZ\alpha/2}\Pi(O)e^{iZ\alpha/2}\;.
 \end{equation}

Only the rotation $\alpha$ and the compression $\lambda$ in the equatorial
plane have any effect on our protocols, and they contribute in a very
straightforward way to the relevant action of $\mathcal{E}$.  The
displacement $t$ and the compression $s$ do not appear in the relevant
action of $\mathcal{E}$, although they can have an indirect effect through
the requirement of complete positivity, which means that their values
constrain the possible value of $\lambda$.

The compression $\lambda$ can come, for example, from random spin flips or
phase changes during transit through the channel.  The rotation by
$\alpha$ is not really a decoherence effect at all; it is an unknown, but
systematic phase shift produced by the quantum channel, which can mimic
the phase offset the Alice and Bob are trying to determine.  It might
arise, for example, from a shift of the energy difference between the two
levels as a qubit traverses the channel.  In a real situation, both
$\lambda$ and $\alpha$ might vary from one use of the channel to the next,
but we assume they are constant for the analysis in the next section.

\section{Effect of decoherence}\label{sec4}

This section contains the main results of the paper.  We analyze the
entangled protocol and the coherent-transport protocol in turn.

\subsection{Entangled protocol}

The density matrix for the $n$-qubit cat state can be written as
\cite{shaji06a}
 \begin{equation}
 \label{eq:ghz}
  \rho_{\rm cat}=
  \frac{1}{2^{n+1}}\left(\bigotimes_{j=1}^n(I_j+Z_j) + \bigotimes_{j=1}^n(I_j-Z_j)
                  + \bigotimes_{j=1}^n(X_j+iY_j)+\bigotimes_{j=1}^n(X_j-iY_j)\right)
  \;.
 \end{equation}
The effect of the channel is studied by analyzing its effect on the four
terms.  Recall that the final measurement by Bob is $O_B=X_B^{\otimes n}$;
since this picks up only the off-diagonal elements in
$\mathcal{E}^{\otimes n}(\rho_{\rm cat})$, we can study the effect of the
map by considering only the last two terms in Eq.~(\ref{eq:ghz}).

Formally, we can write
 \begin{equation}
 \langle O_B\rangle=
 \tr\bigl(X_B^{\otimes n}\mathcal{E}^{\otimes n}(\rho_{\rm cat})\bigr)
 =\tr\bigl(\Pi^{\otimes n}(X_B^{\otimes n})\,\mathcal{E}^{\otimes n}(\rho_{\rm cat})\bigr)
 =\tr\bigl(X_B^{\otimes n}\Pi^{\otimes n}\circ\mathcal{E}^{\otimes n}(\rho_{\rm cat})\bigr)\;.
 \end{equation}
The final form shows that we can discard the first two terms in Eq.~(\ref{eq:ghz}).
The contribution of the third term to the expectation value is
 \begin{equation}
 \label{eq:meas}
  \frac{1}{2^{n+1}}
  \tr\!\left(\bigotimes_{j=1}^n X_{B,j}\Pi\circ\mathcal{E}(X_{A,j}+iY_{A,j})\right)
  =\frac{1}{2^{n+1}}\Bigl(\tr\bigl(X_B\Pi\circ\mathcal{E}(X_A+iY_A)\bigr)\Bigr)^n
  \;.
 \end{equation}
The use of Alice's operators $X_A$ and $Y_A$ here is a consequence of the
fact that Alice prepares the initial cat state.

We can now proceed to calculate the term in large parentheses in
Eq.~(\ref{eq:meas}):
 \begin{eqnarray}
 \tr\bigl(X_B\Pi\circ\mathcal{E}(X_A+iY_A)\bigr)
&=&\lambda\,\tr\bigl(X_B e^{-iZ\alpha/2}(X_A+iY_A)e^{iZ\alpha/2}\bigr) \nonumber\\
&=&\lambda\,\tr\bigl(e^{-iZ\varphi/2}X_A e^{iZ\varphi/2}
e^{-iZ\alpha/2}(X_A+iY_A)e^{iZ\alpha/2}\bigr) \nonumber\\
&=&\lambda\,\tr\bigl(e^{-iZ(\varphi-\alpha)}(I-Z)\bigr) \nonumber\\
&=&2\lambda\langle1|e^{-iZ(\varphi-\alpha)}|1\rangle \nonumber\\
&=&2\lambda\,e^{i(\varphi-\alpha)}\;.
 \end{eqnarray}
Thus the contribution of third term in Eq.~(\ref{eq:meas}) to $\langle
O_B\rangle$ is $\lambda^n e^{in(\varphi-\alpha)}/2$, and the fourth term
contributes the complex conjugate.

All this yields an expectation value
 \begin{equation}
 \label{eq:ent}
\langle O_B \rangle = \langle X_B^{\otimes n}\rangle=
\lambda^n\cos[n(\varphi-\alpha)]\;.
 \end{equation}
The effect of the equatorial plane decoherence is to reduce the fringe
visibility by an exponential factor $\lambda^n$; this exponential
dependence expresses the extreme sensitivity of the entangled protocol to
decoherence in the equatorial plane.  An unknown, systematic phase shift
$\alpha$ is indistinguishable from the phase offset Alice and Bob are
trying to determine and thus limits Bob's ability to determine $\varphi$,
even in the absence of equatorial plane decoherence, i.e., $\lambda=1$.
We set aside this problem for the present, assuming $\alpha=0$, but return
to it after our analysis of decoherence in the coherent-transport
protocol.

The uncertainty in $O_B$,
 \begin{equation}
\label{var1}
\Delta O_B = \sqrt{1-\lambda^{2n}\cos^2(n\varphi)}\;,
 \end{equation}
yields a nominal uncertainty in the estimate of $\varphi$,
 \begin{equation}
\label{eq:var1a}
\Delta\varphi=\frac{\Delta O_B}{|d \avg {O_B}/d\varphi|}
=\frac{\sqrt{1-\lambda^{2n} \cos^2(n\varphi)}}
{n\lambda^n\sin(n\varphi)}\;.
 \end{equation}
The uncertainty~(\ref{eq:var1a}), unlike the $\lambda=1$ limit of
Eq.~(\ref{eq:resolution}), depends on $\varphi$ and, indeed, blows up when
$n(\varphi-\alpha)$ is a multiple of $\pi$, i.e., when one happens to be
at maximum or minimum of the fringe pattern. This is a purely technical
problem, which can be overcome in a variety of ways. For example, in the
extended protocol outlined in Sec.~\ref{sec2B}, in which Bob runs the bare
protocol several times to determine each bit of $T=\varphi/2\pi$, he can
alternate measurements of $X_B^{\otimes n}$ with measurements of
$Y_B^{\otimes n}$, for which $\langle Y_B^{\otimes
n}\rangle=-\lambda^n\sin[n(\varphi-\alpha)]$.  Sampling from fringe
patterns $90^\circ$ out of phase in this way allows Bob always to
determine $\varphi$ with an uncertainty close to optimal, i.e.,
$\Delta\varphi\simeq1/n\lambda^n$.  Comparing this bare sensitivity with
the sensitivity achieved by using $n$ unentangled qubits, $1/\sqrt
n\lambda$, one sees $\lambda$ must be very close to 1 in order to receive
any benefit from entangling substantial numbers of qubits.

\subsection{Coherent-transport protocol}

\begin{figure}[!htb]
\resizebox{4 in}{0.5 in}{\includegraphics{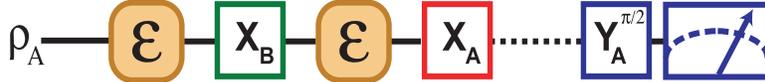}}
\caption{(Color online)
A single ping-pong of the qubit between Alice and Bob in the
coherent-transport protocol, with the decohering CPTP map included in the
two traverses of the quantum channel.}
\label{fig3}
\end{figure}

A single exchange of the qubit between Alice and Bob in the
coherent-transport protocol is shown in Fig.~\ref{fig3}, where
\begin{equation}
\rho_A=|\phi_A\rangle\langle\phi_A|={\frac{1}{2}}(I+X_A)
\end{equation}
is the initial state~(\ref{eq:phiA})
prepared by Alice. The qubit's state on its return to Alice is
 \begin{equation}
X_A\mathcal{E}\bigl(X_B\mathcal{E}(\rho_A)X_B\bigr)X_A=
\mathcal{G}(\rho_A)\;.
 \end{equation}
Here we introduce the overall CPTP map $\mathcal{G}$ for a single
exchange.  What we want to calculate is the expectation value of the
observable $X_A$ measured by Alice after this single exchange.  Using
$\Pi(X_A)=X_A$ and Eq.~(\ref{eq:Pihermiticity}), we can write this
expectation value as
 \begin{equation}
 \langle X_A\rangle=
 \tr\bigl(X_A\mathcal{G}(\rho_A)\bigr)
=\tr\bigl(X_A\Pi\circ\mathcal{G}(\rho_A)\bigr)\;.
 \end{equation}
Now we use Eq.~(\ref{eq:proj}) and the fact that $\Pi$ commutes with
application of $X_A$ and $X_B$ to write
 \begin{equation}
\Pi\circ\mathcal{G}(\rho_A)
=\lambda^2 X_A e^{-iZ\alpha/2}X_B e^{-iZ\alpha/2}\Pi(\rho_A)
e^{iZ\alpha/2}X_B e^{iZ\alpha/2}X_A\;.
 \end{equation}
Equation~(\ref{eq:intro6}) now gives
 \begin{equation}
 X_A e^{-iZ\alpha/2}X_B e^{-iZ\alpha/2}=X_A X_B=e^{iZ\varphi}\;,
 \end{equation}
from which we have
\begin{equation}
\label{eq:PiG}
\Pi\circ\mathcal{G}(\rho_A)=\lambda^2e^{iZ\varphi}\Pi(\rho_A)e^{-iZ\varphi}
={\frac{1}{2}}\lambda^2e^{iZ\varphi}X_Ae^{-iZ\varphi}
={\frac{1}{2}}\lambda^2X_Ae^{-iZ2\varphi}\;.
\end{equation}
Thus the desired expectation value is
\begin{equation}
\langle X_A\rangle=
{\frac{1}{2}}\lambda^2\tr(e^{-iZ2\varphi})=
\lambda^2 \cos(2\varphi)\;.
\end{equation}

These considerations are easily generalized to $r$ exchanges.  The qubit
state after $r$ exchanges is $\mathcal{G}^r(\rho_A)$.  Equation~(\ref{eq:PiG})
generalizes to
\begin{equation}
\Pi\circ\mathcal{G}^r(\rho_A)=\lambda^{2\,r}e^{iZr\varphi}\Pi(\rho_A)e^{-iZr\varphi}
={\frac{1}{2}}\lambda^{2\,r}X_Ae^{-iZ2\,r\varphi}\;,
\end{equation}
which means that the expectation value of a measurement of $X_A$ by Alice
after $r$ exchanges is
\begin{equation}
\langle X_A\rangle=
\tr\bigl(X_A\Pi\circ\mathcal{G}^r(\rho_A)\bigr)
=\lambda^{2\,r}\cos(2\,r\varphi)\;.
\end{equation}
Comparison with the comparable expectation value~(\ref{eq:ent}) for the
entangled protocol shows that the coherent-transport protocol has the same
behavior as the entangled protocol, with $n=2r$, except that the coherent-
transport protocol is insensitive to the systematic channel phase shift
$\alpha$.  In accordance with our discussion of the entangled protocol,
this means that the coherent-transport protocol can determine the phase
offset with uncertainty $\Delta\varphi\simeq1/n\lambda^n$.

The insensitivity of the coherent-transport protocol to $\alpha$ is
noteworthy and deserves discussion.  The insensitivity to $\alpha$ comes
about because Bob's spin flip $X_B$ has the effect that the phase shift
accumulated by the qubit as it traverses the channel from Alice to Bob is
canceled by phase shift on the return leg to Alice.  A slight modification
of the entangled protocol allows it to take advantage of the same effect.
Alice prepares $n/2$ qubits in the cat state.  She sends the qubits to
Bob, who performs his spin flip $X_B$ on each qubit and sends them all
back to Alice.  Alice then measures $X_A^{\otimes n/2}$.  This combination
of the entangled and coherent-transport protocols achieves the same
Heisenberg-limited sensitivity as the coherent-transport protocol and,
like it, is insensitive to an unvarying systematic channel phase shift.

\section{Conclusion}\label{sec5}

In classical clock-synchronization protocols, the uncertainty in the estimate of the time
offset between Alice and Bob goes as $1/\sqrt n$, where $n$ is the number
of uses of a channel between Alice and Bob.  Quantum clock-synchronization
protocols have a better scaling, $1/n$, known as the Heisenberg limit.
Entanglement was originally identified as the resource necessary for a
quantum advantage, but subsequent work showed that coherent transport
without entanglement can achieve the same Heisenberg-limited scaling.  The
communication complexities of the cat-state entangled protocol and the
coherent-transport protocol are identical.  It is natural to ask if this
equivalence is maintained in the presence of decoherence in the quantum
channel between Alice and Bob.  We show in this paper that this is indeed
the case for any channel decoherence that does not itself provide
synchronization information.  The spatial coherence of cat-state
entanglement and the temporal coherence used in coherent transport are
affected in the same way by any such decoherence process.  In analyzing
the effect of decoherence, we found that the cat-state entangled protocol,
unlike the coherent-transport protocol, is sensitive to an unknown,
systematic phase shift induced by the quantum channel, even in the absence
of real decoherence, and we discussed how to eliminate this sensitivity by
combining the entangled protocol with a minimal amount of coherent
transport.

\section*{Acknowledgements}

This work was supported in part by US Office of Naval Research Contract
No.~N00014-03-1-0426. SB acknowledges the support of {\em La Caixa\/} 
fellowship program.

\bibliography{clocks}

\begin{thebibliography}{16}
\expandafter\ifx\csname natexlab\endcsname\relax\def\natexlab#1{#1}\fi
\expandafter\ifx\csname bibnamefont\endcsname\relax
  \def\bibnamefont#1{#1}\fi
\expandafter\ifx\csname bibfnamefont\endcsname\relax
  \def\bibfnamefont#1{#1}\fi
\expandafter\ifx\csname citenamefont\endcsname\relax
  \def\citenamefont#1{#1}\fi
\expandafter\ifx\csname url\endcsname\relax
  \def\url#1{\texttt{#1}}\fi
\expandafter\ifx\csname urlprefix\endcsname\relax\def\urlprefix{URL }\fi
\providecommand{\bibinfo}[2]{#2}
\providecommand{\eprint}[2][]{\url{#2}}

\bibitem[{\citenamefont{Braginsky and Vorontsov}(1975)}]{braginsky75a}
\bibinfo{author}{\bibfnamefont{V.~B.} \bibnamefont{Braginsky}}
  \bibnamefont{and} \bibinfo{author}{\bibfnamefont{Y.~I.}
  \bibnamefont{Vorontsov}}, \bibinfo{journal}{Sov. Phys. Usp.}
  \textbf{\bibinfo{volume}{17}}, \bibinfo{pages}{644} (\bibinfo{year}{1975}).

\bibitem[{\citenamefont{Giovannetti et~al.}(2004)\citenamefont{Giovannetti,
  Lloyd, and Maccone}}]{giovannetti04a}
\bibinfo{author}{\bibfnamefont{V.}~\bibnamefont{Giovannetti}},
  \bibinfo{author}{\bibfnamefont{S.}~\bibnamefont{Lloyd}}, \bibnamefont{and}
  \bibinfo{author}{\bibfnamefont{L.}~\bibnamefont{Maccone}},
  \bibinfo{journal}{Science} \textbf{\bibinfo{volume}{306}},
  \bibinfo{pages}{1330} (\bibinfo{year}{2004}).

\bibitem[{\citenamefont{Caves et~al.}(1980)\citenamefont{Caves, Thorne, Drever,
  Sandberg, and Zimmermann}}]{caves80a}
\bibinfo{author}{\bibfnamefont{C.~M.} \bibnamefont{Caves}},
  \bibinfo{author}{\bibfnamefont{K.~S.} \bibnamefont{Thorne}},
  \bibinfo{author}{\bibfnamefont{R.~W.~P.} \bibnamefont{Drever}},
  \bibinfo{author}{\bibfnamefont{V.~D.} \bibnamefont{Sandberg}},
  \bibnamefont{and}
  \bibinfo{author}{\bibfnamefont{M.}~\bibnamefont{Zimmermann}},
  \bibinfo{journal}{Rev. Mod. Phys.} \textbf{\bibinfo{volume}{52}},
  \bibinfo{pages}{341} (\bibinfo{year}{1980}).

\bibitem[{\citenamefont{Bollinger et~al.}(1996)\citenamefont{Bollinger, Itano,
  Wineland, and Heinzen}}]{bollinger96a}
\bibinfo{author}{\bibfnamefont{J.~J.} \bibnamefont{Bollinger}},
  \bibinfo{author}{\bibfnamefont{W.~M.} \bibnamefont{Itano}},
  \bibinfo{author}{\bibfnamefont{D.~J.} \bibnamefont{Wineland}},
  \bibnamefont{and} \bibinfo{author}{\bibfnamefont{D.~J.}
  \bibnamefont{Heinzen}}, \bibinfo{journal}{Phys. Rev. A}
  \textbf{\bibinfo{volume}{54}}, \bibinfo{pages}{R4649} (\bibinfo{year}{1996}).

\bibitem[{\citenamefont{Huelga et~al.}(1997)\citenamefont{Huelga, Macchiavello,
  Pellizzari, Ekert, Plenio, and Cirac}}]{huelga97a}
\bibinfo{author}{\bibfnamefont{S.~F.} \bibnamefont{Huelga}},
  \bibinfo{author}{\bibfnamefont{C.}~\bibnamefont{Macchiavello}},
  \bibinfo{author}{\bibfnamefont{T.}~\bibnamefont{Pellizzari}},
  \bibinfo{author}{\bibfnamefont{A.~K.} \bibnamefont{Ekert}},
  \bibinfo{author}{\bibfnamefont{M.~B.} \bibnamefont{Plenio}},
  \bibnamefont{and} \bibinfo{author}{\bibfnamefont{J.~I.} \bibnamefont{Cirac}},
  \bibinfo{journal}{Phys. Rev. Lett.} \textbf{\bibinfo{volume}{79}},
  \bibinfo{pages}{3865} (\bibinfo{year}{1997}).

\bibitem[{\citenamefont{Jozsa et~al.}(2000)\citenamefont{Jozsa, Abrams,
  Dowling, and Williams}}]{jozsa00a}
\bibinfo{author}{\bibfnamefont{R.}~\bibnamefont{Jozsa}},
  \bibinfo{author}{\bibfnamefont{D.~S.} \bibnamefont{Abrams}},
  \bibinfo{author}{\bibfnamefont{J.~P.} \bibnamefont{Dowling}},
  \bibnamefont{and} \bibinfo{author}{\bibfnamefont{C.~P.}
  \bibnamefont{Williams}}, \bibinfo{journal}{Phys. Rev. Lett.}
  \textbf{\bibinfo{volume}{85}}, \bibinfo{pages}{2010} (\bibinfo{year}{2000}).

\bibitem[{\citenamefont{Chuang}(2000)}]{chuang00a}
\bibinfo{author}{\bibfnamefont{I.~L.} \bibnamefont{Chuang}},
  \bibinfo{journal}{Phys. Rev. Lett.} \textbf{\bibinfo{volume}{85}},
  \bibinfo{pages}{2006} (\bibinfo{year}{2000}).

\bibitem[{\citenamefont{Preskill}(2000)}]{preskill00a}
\bibinfo{author}{\bibfnamefont{J.}~\bibnamefont{Preskill}},
  \bibinfo{journal}{e-print quant-ph/001098}  (\bibinfo{year}{2000}).

\bibitem[{\citenamefont{Yurtsever and Dowling}(2002)}]{yurtsever02a}
\bibinfo{author}{\bibfnamefont{U.}~\bibnamefont{Yurtsever}} \bibnamefont{and}
  \bibinfo{author}{\bibfnamefont{J.~P.} \bibnamefont{Dowling}},
  \bibinfo{journal}{Phys. Rev. A} \textbf{\bibinfo{volume}{65}},
  \bibinfo{pages}{052317} (\bibinfo{year}{2002}).

\bibitem[{\citenamefont{Revzen and Mann}(2003)}]{revzen03a}
\bibinfo{author}{\bibfnamefont{M.}~\bibnamefont{Revzen}} \bibnamefont{and}
  \bibinfo{author}{\bibfnamefont{A.}~\bibnamefont{Mann}},
  \bibinfo{journal}{Phys. Lett. A} \textbf{\bibinfo{volume}{312}},
  \bibinfo{pages}{11} (\bibinfo{year}{2003}).

\bibitem[{\citenamefont{de~Burgh and Bartlett}(2005)}]{burgh05a}
\bibinfo{author}{\bibfnamefont{M.}~\bibnamefont{de~Burgh}} \bibnamefont{and}
  \bibinfo{author}{\bibfnamefont{S.~D.} \bibnamefont{Bartlett}},
  \bibinfo{journal}{Phys. Rev. A} \textbf{\bibinfo{volume}{72}},
  \bibinfo{pages}{042301} (\bibinfo{year}{2005}).

\bibitem[{\citenamefont{Rudolph and Grover}(2003)}]{rudolph03a}
\bibinfo{author}{\bibfnamefont{T.}~\bibnamefont{Rudolph}} \bibnamefont{and}
  \bibinfo{author}{\bibfnamefont{L.}~\bibnamefont{Grover}},
  \bibinfo{journal}{Phys. Rev. Lett.} \textbf{\bibinfo{volume}{91}},
  \bibinfo{pages}{217905} (\bibinfo{year}{2003}).

\bibitem[{\citenamefont{Salecker and Wigner}(1958)}]{salecker58a}
\bibinfo{author}{\bibfnamefont{H.}~\bibnamefont{Salecker}} \bibnamefont{and}
  \bibinfo{author}{\bibfnamefont{E.~P.} \bibnamefont{Wigner}},
  \bibinfo{journal}{Phys. Rev.} \textbf{\bibinfo{volume}{109}},
  \bibinfo{pages}{571} (\bibinfo{year}{1958}).

\bibitem[{\citenamefont{Braunstein and Caves}(1994)}]{braunstein94a}
\bibinfo{author}{\bibfnamefont{S.~L.} \bibnamefont{Braunstein}}
  \bibnamefont{and} \bibinfo{author}{\bibfnamefont{C.~M.} \bibnamefont{Caves}},
  \bibinfo{journal}{Phys. Rev. Lett} \textbf{\bibinfo{volume}{72}},
  \bibinfo{pages}{3439} (\bibinfo{year}{1994}).

\bibitem[{\citenamefont{Nielsen and Chuang}(2000)}]{nielsen00a}
\bibinfo{author}{\bibfnamefont{M.~A.} \bibnamefont{Nielsen}} \bibnamefont{and}
  \bibinfo{author}{\bibfnamefont{I.~L.} \bibnamefont{Chuang}},
  \emph{\bibinfo{title}{Quantum Computation and Quantum Information}}
  (\bibinfo{publisher}{Cambridge University Press},
  \bibinfo{address}{Cambridge}, \bibinfo{year}{2000}).

\bibitem[{\citenamefont{Shaji and Caves}(2006)}]{shaji06a}
\bibinfo{author}{\bibfnamefont{A.}~\bibnamefont{Shaji}} \bibnamefont{and}
  \bibinfo{author}{\bibfnamefont{C.~M.} \bibnamefont{Caves}}
  (\bibinfo{year}{2006}), \bibinfo{note}{to be published}.

\end{thebibliography}

\end{document}